\documentclass[twocolumn,aps,superscriptaddress,showpacs,nofootinbib,floatfix]{revtex4}

\usepackage{epsfig,bm,feynmf}

\usepackage{graphics}

\usepackage[normalem]{ulem}  
\usepackage[dvips]{color} 

\renewcommand\sout{\bgroup \color{red} \ULdepth=-.5ex \ULset}

\begin{document}



\title{Bottomonia suppression in heavy-ion collisions}


\author{Taesoo Song}\email{songtsoo@yonsei.ac.kr}
\affiliation{Cyclotron Institute, Texas A$\&$M University, College Station, TX 77843-3366, USA}
\author{Kyong Chol Han}\email{khan@comp.tamu.edu}
\affiliation{Cyclotron Institute and Department of Physics and Astronomy, Texas A$\&$M University, College Station, TX 77843-3366, USA}
\author{Che Ming Ko}\email{ko@comp.tamu.edu}
\affiliation{Cyclotron Institute and Department of Physics and Astronomy, Texas A$\&$M University, College Station, TX 77843-3366, USA}


\begin{abstract}
Using the two-component model that includes both initial production from nucleon-nucleon hard scattering and regeneration from produced quark-gluon plasma (QGP), we study the effect of medium modifications of the binding energies and radii of bottomonia on their production in heavy-ion collisions.  We find that the contribution to bottomonia production from regeneration is small and the inclusion of medium effects is generally helpful for understanding the observed suppression of bottomonia production in experiments carried out at both the Relativistic Heavy Ion Collider (RHIC) and the Large Hadron Collider (LHC).
\end{abstract}

\pacs{25:75.Cj} \keywords{}

\maketitle


\section{introduction}

Since the suggestion by Matsui and Satz~\cite{Matsui:1986dk} that suppressed production of $J/\psi$ in relativistic heavy-ion collisions could be a signature for produced QGP, there have been many experimental~\cite{Alessandro:2004ap,Adare:2006ns} and theoretical studies~\cite{Vogt:1999cu,Zhang:2000nc,Zhang:2002ug,Zhao:2007hh,Yan:2006ve,Song:2010ix,Song:2010er,Song:2011xi} on this very interesting phenomenon; see, e.g., Refs. \cite{Rapp:2008tf,Andronic:2006ky} for a recent review. Although their original idea was that the screening of color charges in the QGP would prohibit charm and anticharm quarks from forming the $J/\psi$ and thus suppress its production, lattice QCD calculations of the $J/\psi$ spectral function showed, on the other hand, that the $J/\psi$ could survive inside QGP up to the so-called dissociation temperature~\cite{Hatsuda04,Datta04}. However, the dissociation temperature depends on how the $J/\psi$ spectral function is extracted from lattice data and it is not yet clear if it is far from the critical temperature~\cite{Mocsy:2007yj}. Studying $J/\psi$ suppression in relativistic heavy-ion collisions can thus provide not only a signature for the QGP but also a probe of its properties \cite{Alessandro:2004ap,Adare:2006ns,:2010px,Dahms:2011gn}. A quantitative study of $J/\psi$ production in heavy-ion collisions is, however, complicated by their absorption in the initial cold nuclear matter and regeneration in the QGP. Since the $\Upsilon$ is a more strongly bound state of bottom and antibottom quarks than the $J/\psi$, its production in heavy-ion collisions is expected to be less affected by initial cold nuclear matter effects. Furthermore, the much smaller number of bottom quarks than charm quarks that is produced in heavy-ion collisions makes the contribution of regeneration from the QGP to $\Upsilon$ production also less important. Therefore, studying $\Upsilon$ production in heavy-ion collisions would provide a cleaner probe of the properties of QGP and also the in-medium properties of bottomonia~\cite{Grandchamp:2005yw,Liu:2010ej}. Recently, the nuclear modification factor $R_{AA}$ of the sum of bottomonia $\Upsilon$(1S), $\Upsilon$(2S) and $\Upsilon$(3S), defined by their yields relative to those from p+p collisions multiplied by the number of initial binary collisions, in Au+Au collisions at $\sqrt{s_{NN}}=200$ GeV was measured by the STAR Collaboration at RHIC \cite{Rosi}, while that of $\Upsilon$(1S) \cite{cms} and the relative suppression of $\Upsilon$(2S) and $\Upsilon$(3S) to $\Upsilon$(1S) \cite{Chatrchyan:2011pe} in Pb+Pb collisions at $\sqrt{s_{NN}}=2.76$ TeV were measured by the CMS Collaboration at LHC. In both experiments, the measured $R_{AA}$ was seen to decrease with increasing centrality of collisions. In the present study, we use the two-component model~\cite{Song:2010ix,Song:2010er,Song:2011xi}, which was previously used for studying $J/\psi$ production in heavy-ion collisions, to show that these experimental results allow us to obtain useful information on the properties of bottomonia in QGP.

The paper is organized as follows. We first briefly review in Sec.~\ref{two} the two-component model for bottomonia production in heavy-ion collisions. We then compare in Sec.~\ref{data} the calculated nuclear modification factors of bottomonium from the model with experimental data. Finally, a conclusion is given in Sec. \ref{conclusion}.

\section{the two-component model}\label{two}

In the two-component model, bottomonia are produced from both initial nucleon-nucleon hard scattering and regeneration in the produced QGP. The numbers of initially produced bottomonia are proportional to the number of binary collisions among nucleons in the two colliding nuclei. Whether these bottomonia can survive after collisions depends on many effects from both the initial cold nuclear matter and the final hot partonic and hadronic matters. For the initial cold nuclear matter effect, the shadowing and the nuclear absorption effect are included. The shadowing is the modification of the parton distribution in a nucleus from that in a single nucleon. In this study, we use the EPS09 package \cite{Eskola:2009uj} and assume that the effect is proportional to the path length of a parton in nucleus \cite{Vogt:2004dh,Song:2011xi}. Using the experimental data on the average transverse momentum of $\Upsilon$ \cite{Matthew,cms}, the ratios of gluon distribution function in a nucleus to that in a single nucleon at $x=m_T/\sqrt{s_{NN}}$ are 1.12 and 0.92 for collisions at RHIC and LHC energies, respectively, with the ratio larger than one called the anti-shadowing. For the nuclear absorption cross section of bottomonium in cold nuclear matter, its value is about 4.3 mb at RHIC based on the recently measured  nuclear modification factor $R_{\rm dAu}=0.78\pm 0.28$ of midrapidity bottomonium in d+Au collisions at RHIC~\cite{Reed:2011zz}. This value is larger than the value of 2.8 mb for the nuclear absorption cross section of charmonium in heavy-ion collisions at RHIC~\cite{Adare:2007gn}. The nuclear absorption cross section of bottomonia at LHC is, however, not known empirically. Since the nuclear cross section of bottomonium is expected to decrease with increasing collisions energy, we take the value 4.3 mb as an upper limit at LHC and consider the two cases of 0 and 4.3 mb. For thermal dissociation of bottomonia, it includes the initial dissociation in the produced QGP of temperatures above their dissociation temperatures and the subsequent dissociation by quarks and gluons in the QGP and hadrons in the hadronic matter (HG).

For the local initial temperature, we estimate it from the local entropy density
$ds/d\eta=C[(1-\alpha)n_{\rm part}/2+\alpha~n_{\rm coll}]$ \cite{Kharzeev:2000ph} in terms of the number densities of participants and binary collisions $n_{\rm part(coll)}=\Delta N_{\rm part(coll)}/(\tau_0\Delta x \Delta y)$ with $\Delta N_{\rm part(coll)}$ being the number of participants (binary collisions) in the volume $\tau_0\Delta x\Delta y$ of the transverse area $\Delta x\Delta y$ and $\tau_0$ being the initial thermalization time. For the value of $\tau_0$, we use 0.9 and 1.05 fm/c from the viscous hydrodynamics \cite{Song:2011qa} for Au+Au collisions at $\sqrt{s_{NN}}=200$ GeV at RHIC and Pb+Pb collisions at $\sqrt{s_{NN}}=2.76$ TeV at LHC, respectively. The parameters $C$ and $\alpha$ are, respectively, 0.11 and 18.7 for RHIC and 0.15 and 27.0 for LHC from fitting measured charged particle multiplicities~\cite{Song:2011xi}. With the quasiparticle model of three flavors for the equation of state of QGP and the resonance gas model for that of HG~\cite{Levai:1997yx,Song:2010ix}, the initial maximum temperatures are 324 and 390 MeV and mean temperatures are 269 and 311 MeV for central collisions at RHIC and LHC, respectively. After initial thermalization, a schematic viscous hydrodynamics \cite{Song:2010fk,Song:2011xi} is then used to describe the expansion of the hot dense matter with the specific shear viscosity $\eta/s$ taken to be 0.16 and 0.2 for the QGP at RHIC and LHC \cite{Song:2011qa}, respectively, and 0.8 for the HG \cite{Demir:2008tr}.

The thermal dissociation by partons in the QGP or hadrons in the HG of bottomonia that have survived from initial dissociations is described by the rate equation for the number $N_i$ of bottomonia of type $i$
\begin{eqnarray}
\frac{dN_i}{d\tau}=-\Gamma_i(N_i-N_i^{\rm eq}),
\label{rate}
\end{eqnarray}
where $\tau$ is the longitudinal proper time and $\Gamma_i$ is the thermal decay width of bottomonia. The number of equilibrated bottomonia of type $i$ is given by $N_i^{\rm eq}=\gamma^2 R~ n_i ~f V\theta(T_i-T)$, where $n_i$ is its number density in grandcanonical ensemble; $f$ is the fraction of QGP in the mixed phase and is 1 in the QGP; and $\theta(T_i-T)$ is the step function with $T_i$ being the dissociation temperature of bottomonia of type $i$. The chemical and kinetic off-equilibrium of bottom quarks in the QGP is included through their fugacity $\gamma$ and relaxation factor $R$, respectively, with the former obtained from the conservation of bottom flavor \cite{Song:2011xi} and the latter defined as $R(\tau)=1-\exp[-(\tau-\tau_0)/\tau_{\rm eq}]$ where $\tau_{\rm eq}$ is the relaxation time of bottom quarks. With $\tau_{\rm eq}\sim m_Q$ \cite{Moore:2004tg} and $\tau_{\rm eq}=$4 fm/c for charm quarks \cite{Song:2011xi}, the relaxation time for bottom quarks is about 14 fm/c. We note that the second term on the right hand of Eq.(\ref{rate}) takes into account the regeneration of bottomonia from bottom and antibottom quarks in the QGP.

\begin{figure}[h]
\centerline{
\includegraphics[width=6.8cm]{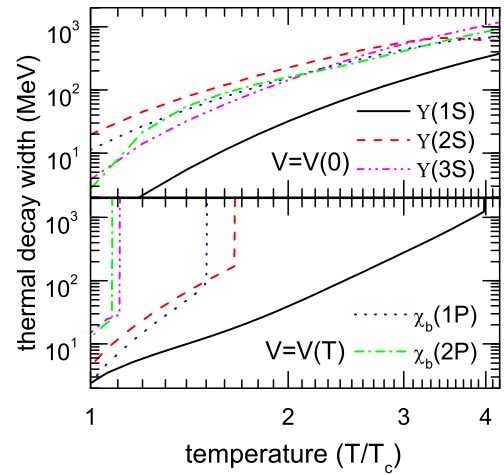}}
\caption{Thermal decay widths of bottomonia as functions of the temperature of QGP without (upper panel) and with (lower panel) medium effects.}
\label{widths}
\end{figure}

For the thermal decay widths of bottomonia in QGP, we calculate them up to the next-to-leading order (NLO) in perturbative QCD (pQCD) \cite{Song:2005yd}. While in the leading order (LO) a bottomonium is dissociated by absorbing a thermal gluon, in the NLO it is dissociated by the gluon emitted from a quark or gluon in the QGP. The squared invariant amplitudes for these processes are the same as those given in Ref. \cite{Song:2011xi} for charmonia except the heavy quark mass and can be found in Ref. \cite{Song:2011xi}. In term of the resulting bottomonium dissociation cross section $\sigma^{\rm diss}$,
the thermal decay width of a bottomonium is given by
\begin{eqnarray}
\Gamma(T)=\sum_i \int\frac{d^3k}{(2\pi)^3}v_{\rm rel}(k)n_i(k,T) \sigma_i^{\rm diss}(k,T),
\label{width}
\end{eqnarray}
where $i$ denotes the quarks and gluons in the QGP; $n_i$ is the number density of parton species $i$ in grand canonical ensemble; and $v_{\rm rel}$ is the relative velocity between the scattering bottomonium and parton. For the thermal width in the mixed phase, it is taken to be a linear combination of those in the QGP and the HG, i.e., $\Gamma(T_c)=f~\Gamma^{\rm QGP}(T_c)+(1-f)\Gamma^{\rm HG}(T_c)$, where $f$ is the fraction of QGP in the mixed phase. For the dissociation cross section of $\Upsilon$(1S) in the hadron gas (HG), we use the factorization formula \cite{Song:2005yd} and assume that those of excited bottomonia are proportional to their squared radii.  In the upper panel of Fig.~\ref{widths}, we show the thermal decay widths of bottomonia in QGP calculated with their masses and radii in free space, i.e., 9.5 GeV and 0.23 fm, 9.97 GeV and 0.59 fm, 10.19 GeV and 0.95 fm, 9.9 GeV and 0.46 fm, and 10.25 GeV and 0.82 fm for $\Upsilon(1S)$, $\Upsilon(2S)$, $\Upsilon(3S)$, $\chi_{\rm b}(1P)$, and $\chi_{\rm b}(2P)$, respectively, obtained from the Cornell potential with the vacuum screening mass $\mu=$0.18 GeV~\cite{Karsch:1987pv}. These thermal decay widths of bottomonia are appreciable and increase with increasing temperature. We note that the thermal decay widths of bottomonia in HG are significantly smaller than those in QGP.

\begin{figure}[h]
\centerline{
\includegraphics[width=6.8cm]{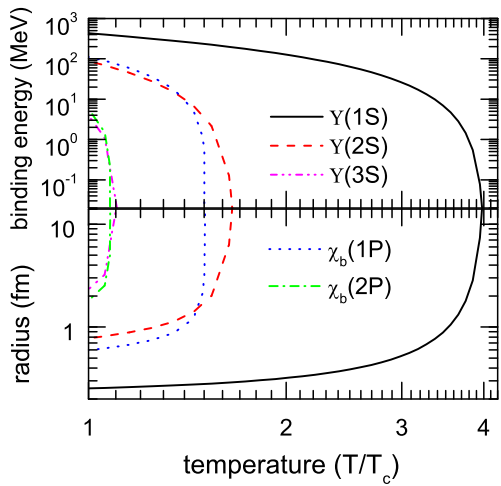}}
\caption{Binding energies (upper panel) and radii (lower panel) of bottomonia as functions of the temperature of QGP.}
\label{properties}
\end{figure}

To include medium effects on the properties of bottomonia in QGP, we consider the modification of the potential energy between bottom and antibottom quarks due to the Debye screening of color charges. This is achieved by using the screened Cornell potential \cite{Karsch:1987pv}:
\begin{eqnarray}
V(r,T)=\frac{\sigma}{\mu(T)}\bigg[1-e^{-\mu(T) r}\bigg]-\frac{\alpha}{r}e^{-\mu(T) r}
\label{Cornell}
\end{eqnarray}
with $\sigma=0.192~{\rm GeV^2}$ and $\alpha=0.471$. The screening mass $\mu(T)$ depends on temperature, and we use the one given in pQCD, i.e., $\mu(T)=\sqrt{N_c/3+N_f/6}~gT$, where $N_c$ and $N_f$ are numbers of colors and light quark flavors, respectively. We note that compared to results from the lattice QCD calculations for a heavy quark and antiquark pair in the QGP \cite{Kaczmarek:1900zz}, this potential is closer to their internal energy at high temperature but to their free energy at low temperature.
The screened Cornell potential thus interpolates smoothly the expected temperature-dependent potential between heavy quark and antiquark~\cite{Wong:2004zr}. The binding energies and radii of bottomonia in the QGP can be obtained by solving the resulting Schr\"odinger equation for the bottom and anti-bottom quark pair. Taking their masses to be $m_b=$ 4.746 GeV \cite{Karsch:1987pv} and the QCD coupling constant $g=1.87$ as in our previous study of $J/\psi$ suppression in heavy-ion collisions \cite{Song:2011xi}, the results are shown in the upper and lower panels of Fig. \ref{properties}. It is seen that the dissociation temperatures of $\Upsilon$(1S), $\Upsilon$(2S), $\Upsilon$(3S), $\chi_b$(1P) and $\chi_b$(2P) are 4.0, 1.67, 1.12, 1.51 and 1.09 $T_c$, respectively, and their radii increase with increasing temperature.
The thermal decay widths of bottomonia obtained with their in-medium binding energies and radii
are shown in the lower panel of Fig.~\ref{widths}, and they increase with temperature and diverge at their dissociation temperatures. Compared to those obtained with binding energies and radii in free space, medium effects enhance the widths significantly.
We note that the inclusion of thermal decay widths of heavy quarkonia effectively takes into account the imaginary part of the potential energy between heavy quark and antiquark at finite temperature \cite{Laine:2006ns,Beraudo:2007ky}.

\section{results}\label{data}

\begin{figure}[h]
\centerline{
\includegraphics[width=6.8cm]{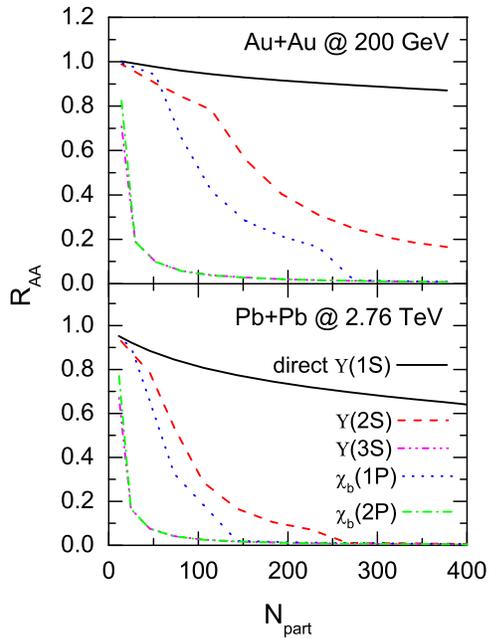}}
\caption{Nuclear modification factor $R_{\rm AA}$ of bottomonia in Au+Au collisions at $\sqrt{s_{NN}}=200$ GeV at RHIC (upper panel) and in Pb+Pb collisions at $\sqrt{s_{NN}}=2.76$ GeV at LHC (lower panel) without the inclusion of the cold nuclear matter effect}
\label{components}
\end{figure}

To calculate the nuclear modification factor of bottomonia in heavy-ion collisions requires information on their numbers produced in p+p collisions at same energy. Since the numbers of bottomonia produced in p+p collisions at LHC are not available, we use in the present study those of $\Upsilon$(1S), $\Upsilon$(2S) and $\Upsilon$(3S) from the experimental data in p+$\bar{p}$ collisions at $\sqrt{s_{NN}}=1.8$ TeV measured by the CDF Collaboration at the FermiLab \cite{Abe:1995an}, and those of $\chi_b$(1P) and $\chi_b$(2P) from their contributions to $\Upsilon$(1S) \cite{Affolder:1999wm} based on the branching ratios of about 0.24 and 0.13, respectively. For p+p collisions at RHIC, the numbers of bottomonia were not individually measured, so we use $\sum_{n=1\sim 3} B(nS)\times d\sigma/dy|_{y=0}(nS)=114 ~{\rm pb}$ \cite{Reed:2011zz}, where $B(nS)$ and $d\sigma/dy|_{y=0}(nS)$ are, respectively, the branching ratio and differential cross section in rapidity for $\Upsilon(nS)$, and assume their relative abundances are the same as those at the LHC. For the initial number of bottom quark pairs in determining the fugacity of bottom flavor, it is obtained from $d\sigma_{b\bar{b}}^{pp}/dy|_{y=0}=1.34 ~{\rm \mu b}$ for RHIC \cite{Morino:2008nc} and $\sigma_{b\bar{b}}^{pp}=17.6 ~{\rm \mu b}$ in the rapidity range $|y|<0.6$ from the CDF Collaboration \cite{Acosta:2004yw} for LHC. Also, we need the contribution of excited bottomonia to $\Upsilon(1S)$ in p+p collisions at same energy. Since this has not been measured, we use the information obtained from p+p collisions at $\sqrt{s_{NN}}=1.97$ TeV by the CDF Collaboration at the FermiLab~\cite{Affolder:1999wm} as they are known to be essentially independent of the collision energy \cite{Abe:1995an}, i.e., the contributions from $\chi_b(1P)$, $\chi_b(2P)$, $\Upsilon(2S)$ and $\Upsilon(3S)$ to $\Upsilon(1s)$ are taken to be 27.1, 10.5, 10.7 and 0.8 \%, respectively. The resulting nuclear modification factors ($R_{\rm AA}$) of bottomonia obtained without the cold nuclear effect in Au+Au collisions at $\sqrt{s_{NN}}=200$ GeV at RHIC and in Pb+Pb collisions at $\sqrt{s_{NN}}=2.76$ GeV at LHC are shown in Fig.~\ref{components}. It is seen that the $R_{\rm AA}$ of directly produced $\Upsilon$(1S) is close to one even in central collisions while those of excited bottomnoia are small.

\begin{figure}[h]
\centerline{
\includegraphics[width=6.8cm]{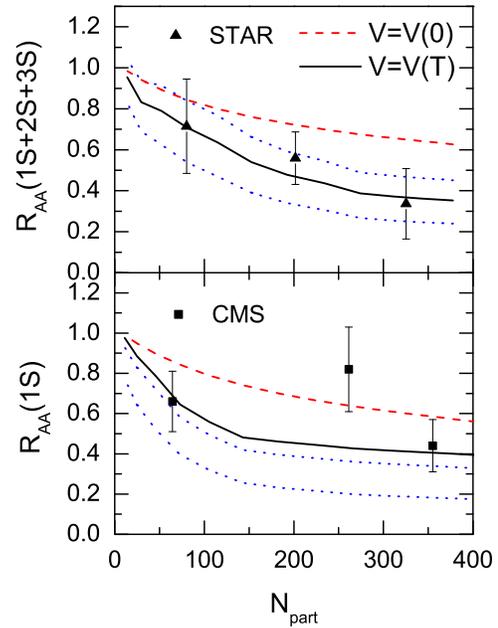}}
\caption{Nuclear modification factor $R_{AA}$ of the sum of $\Upsilon$(1S), $\Upsilon$(2S) and $\Upsilon$(3S) in Au+Au collisions at $\sqrt{s_{NN}}=200$ GeV at RHIC (upper panel) and that of $\Upsilon$(1S) in Pb+Pb collisions at $\sqrt{s_{NN}}=2.76$ TeV at LHC (lower panel) as functions of the participant number. Solid and dashed lines are, respectively, results with and without medium effects on bottomonia. While the upper dotted line is the result including also the shadowing effect, the lower dotted line further includes the nuclear absorption with a cross section of 4.3 mb. Experimental data are taken from Refs.~\cite{Rosi,cms}.}
\label{Raa}
\end{figure}

In the upper panel of Fig.~\ref{Raa}, the calculated $R_{\rm AA}$ of the sum of $\Upsilon$(1S), $\Upsilon$(2S) and $\Upsilon$(3S)
in Au+Au collisions at $\sqrt{s_{NN}}=200$ GeV at RHIC as a function of the participant number is shown and compared with the experimental data from the STAR Collaboration~\cite{Rosi}. The solid line is obtained without the cold nuclear matter effect, while the upper dotted line includes only shadowing effect and the lower dotted line includes both the shadowing and the nuclear absorption effect. It is seen that because of the large experimental errors, results from our model with and without the cold nuclear matter effect all can describe the data for all centralities from the RHIC.

For Pb+Pb collisions at $\sqrt{s_{NN}}=2.76$ TeV at LHC, only the $R_{\rm AA}$ of $\Upsilon(1S)$ has been measured by the CMS Collaboration~\cite{cms}. Our results for the $R_{\rm AA}$ of $\Upsilon$(1S) without the cold nuclear matter effect is shown by the solid line in the lower panel of Fig.~\ref{Raa}. Compared with that measured by the CMS Collaboration \cite{cms}, these results agree with the data for both peripheral (20-100\%) and most central (0-10\%) collisions. For midcentral (10-20\%) collisions, our model significantly underestimates the measured $R_{AA}$. Similar results for LHC were also obtained in Ref.~\cite{Strickland:2011mw} based on the bottom and antibottom quark potential that was taken to be their internal energy from the lattice QCD. Also shown in the lower panel of Fig.~\ref{Raa} are results obtained by including the shadowing effect (upper dotted line) and also the nuclear absorption with a cross section of 4.3 mb (lower dotted line). It is seen that the inclusion of the cold nuclear matter effect, particular the nuclear absorption, leads to too small bottomonia $R_{AA}$ compared with the experimental data. We note that most of the suppression of $\Upsilon$(1S) comes from those of its excited states, as seen from the results shown in Fig.~\ref{components}. Also, the contribution from regeneration the the $R_{AA}$ of bottomonia is less than 1\% at both RHIC and LHC and for all centralities as a result of the small number of bottom quarks and the much longer bottom quark relaxation time than the lifetime of produced QGP. For comparison, the $R_{\rm AA}$'s for the case without medium effects on bottomonia are shown by dashed lines in Fig.~\ref{Raa}, and they are larger than those with medium effects as expected. In this case, the calculated $R_{AA}$ of $\Upsilon(1S)$ is large compared to the experimental data from RHIC, particular for more central collisions. This is also the case for heavy ion collisions at the LHC except for midcentral collisions where the result obtained without medium effects can better describe the experimental data. It is not clear if this indicates a change of the bottomonia suppression mechanism in midcentral collisions. Improved experimental data are essential for resolving this puzzle.

\section{conclusion}\label{conclusion}

In conclusion, using the two-component model \cite{Song:2011xi} that includes both initial production from nucleon-nucleon hard scattering and regeneration from produced quark-gluon plasma, we have studied bottomonia production in heavy-ion collisions at RHIC and LHC by including the medium effects on the thermal properties of bottomonia and their dissociation cross sections. With the expansion dynamics of produced hot dense matter described by a schematic viscous hydrodynamics and including the thermal dissociation of bottomonia as well as the regeneration of bottomonia by using a rate equation, our model describes successfully the experimental data from RHIC and reasonably those from LHC on bottomonia suppression. Our results indicate that the contribution of regenerated bottomonia is small. We have also studied the cold nuclear matter effect due to the shadowing or anti-shadowing of the parton distribution function in the nucleus and the nuclear absorption of bottomonia on their $R_{AA}$. Although the anti-shadowing effect at RHIC enhances the $R_{AA}$, the nuclear absorption reduces it. Present experimental data at RHIC are, however, not accurate enough to differentiate these effects. For the $R_{AA}$ of bottomonia in heavy-ion collisions at LHC, the experimental data are found to be better described by the absence of nuclear absorption, consistent with the expected decreasing effect of nuclear absorption with increasing collision energy. Furthermore, the present study shows that the inclusion of medium effects on bottomonia is essential for describing the experimental observations at RHIC aw well as at LHC except for midcentral collisions where results without medium effects can better describe the data. More accurate data from future experiments are needed to obtain more definitive
information on the properties of bottomonia in the QGP.

\section*{Acknowledgements}

This work was supported in part by the U.S. National Science
Foundation under Grant Nos. PHY-0758115 and PHY-1068572, the US Department of Energy
under Contract No. DE-FG02-10ER41682, and the Welch Foundation under Grant No. A-1358.


\end{document}